\documentclass[aps,prb,twocolumn,amsmath,amssymb,nofootinbib,floatfix,superscriptaddress]{revtex4-1}

\usepackage{amsmath,braket}
\usepackage{amssymb}
\usepackage{amsthm}
\usepackage[dvipdf]{color}
\usepackage{graphicx}
\usepackage{dcolumn}
\usepackage{bm,subfigure} 
\usepackage{xcolor,colortbl}
\usepackage[normalem]{ulem}

\newtheorem{theorem}{Theorem}

\usepackage{enumitem}

\newtheorem{lemma}[theorem]{Lemma}

\begin{document}

\title{Assur graphs, marginally jammed packings, and reconfigurable metamaterials}

  \author{Jose Ortiz}
  \author{Ethan Stanifer}
  \author{Xiaoming Mao}
 \affiliation{
 Department of Physics,
  University of Michigan, Ann Arbor, 
 MI 48109-1040, USA
 }

\begin{abstract} 
Isostatic frames are mechanical networks that are simultaneously rigid and free of self-stress states, and is a powerful concept in understanding phase transitions in soft matter and designing of mechanical metamaterials.  Here we analyze substructures of isostatic frames, by generalizing ``Assur graphs'' to the torus and examine them in physical systems.  We show that the contact network of marginally jammed packings approach torus Assur graphs in the thermodynamic limit, and demostrate how Assur graphs offer a new design principle for  mechanical metamaterials in which motion and stress can propagate in reconfigurable pathways, while rigidity  of the entire structure is maintained.
\end{abstract}
\maketitle

\emph{Introduction.---}
The concept of rigidity is central to both soft matter physics, from gelation~\cite{Zhang2019,colombo2014stress}, jamming~\cite{OHern2003} to mechanical transitions in biological tissues~\cite{bi2015density}, and the design of mechanical metamaterials where deformation and stress responses are programmed via geometry and topology~\cite{Lubensky2015,Mao2018}. 
Given the tensorial nature of rigidity (as opposed to connectivity or conductivity which are scalars), rigidity transitions can take many forms characterized by distinct universality classes.  Among them,  
the special point of ``isostaticity'', where the numbers of nontrivial zero modes (ZMs, normal modes of zero energy) and states of self stress (SSSs, force-balanced stress eigenstates) both vanish, characterizes a type of mechanical critical point where the whole system is coordinated in a unique way where all degrees of freedom are marginally constrained~\cite{laman1970graphs,asimow1978rigidity,alexander1998amorphous}.  It is known that the jamming transition of athermal frictionless repulsive disks (or spheres in 3D) occurs at the isostatic point~\cite{OHern2003,van2009jamming,liu2010jamming,goodrich2016scaling}, whereas many other rigidity transitions don't~\cite{jacobs1995generic,donev2004linear,Mailman2009,broedersz2011criticality,Ellenbroek_2011,bi2011jamming,Yan2014,rigidity_loss_three,Henkes2016,behringer2018physics,Zhang2019}. 

\begin{figure*}
    \centering
    \includegraphics[width=.9\textwidth]{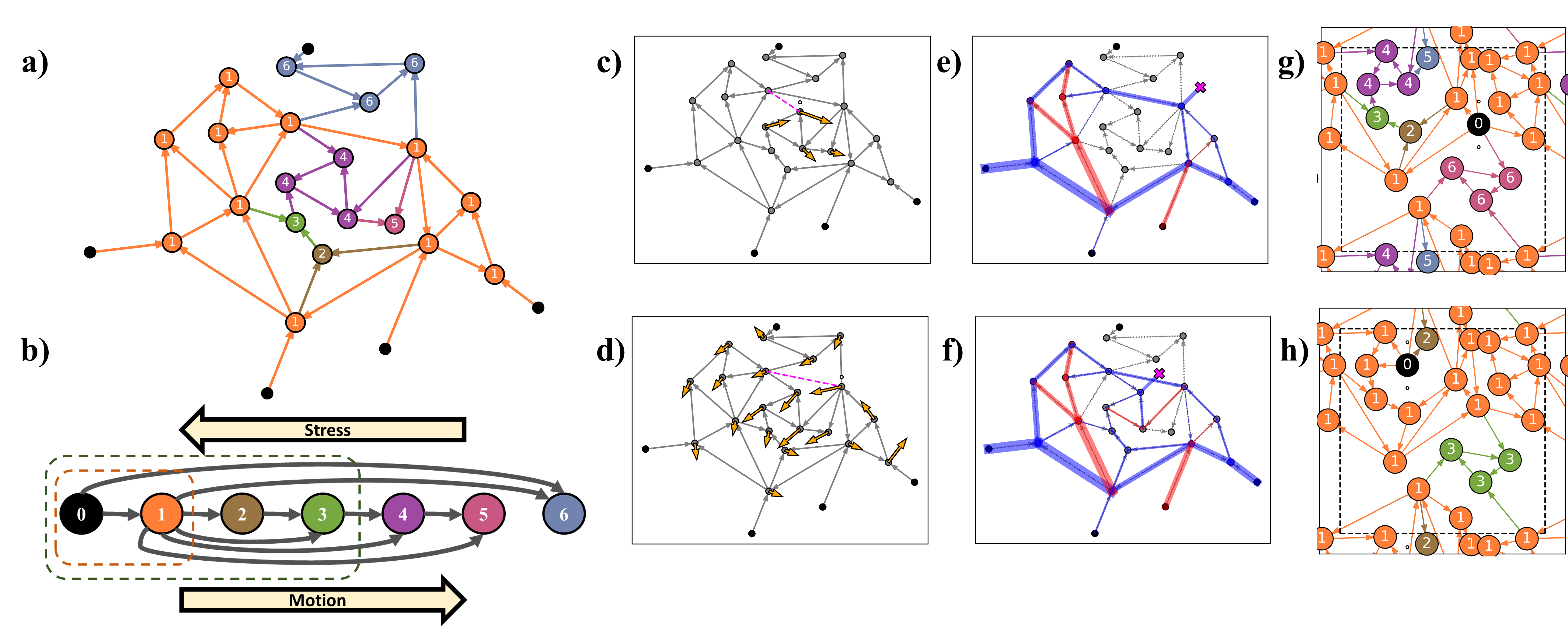}
    \caption{ Assur decomposition for pinned and torus isostatic graphs. (a) A pinned isostatic graph with its 6 Assur components identified. (b) Partial order of the Assur components of the graph in (a), with two isostatic subgraphs identified with boxes and the directions for the propagation of motion and stress marked. 
    In the following panels the pruned edges are marked by  dashed magenta lines, external forces represented as the addition of a redundant edge to ground (magenta cross). The resulting ZM is plotted in orange arrows (with the displacement amplitude proportional to the length of the line excluding the arrowheads) and the resulting stress in red (tension) and blue (compression).
    (c) Pruning an edge in component 4 results in a ZM that only moves component 4 and its descendants. (d) Pruning an edge in component 1 results in a full ZM that moves all vertices.  e) External force on a vertex of component 4 stresses  component 4 and its ancestors. f) External force on component 1 stresses only component 1. (g,h) Assur decompositions of a torus isostatic graph, where different choices of the ground (black ``0'' vertex with 2 pebbles) led to different decompositions. PBC boxes are shown as black dashed squares. This ambiguity is avoided in the torus PGA algorithm we use (see SM). 
    }
    \label{fig:AssurDecomposition}
\end{figure*}

It is known to the mechanics and mathematics literature that pinned isostatic frames can have substructures, called ``Assur graphs'', which define minimal components that determine the propagation of motion and stress (Fig.~\ref{fig:AssurDecomposition})~\cite{assur1952issledovanie,DigraphsDecomp,pinned_pebble_game,uniqueenginprops,PENNE199437}. Interestingly, recently it was found that marginally jammed packings (MJPs) of frictionless repulsive disks are not only isostatic but this isostaticity is ``global''~\cite{rigidity_loss_three},
whereas computational modular representations have been proposed to characterize mechanical networks with floppy motion~\cite{chen2022modular}.

Here we generalize  Assur decomposition from the case of pinned graphs to graphs on the torus, and apply it to MJPs under periodic boundary conditions (PBC) to reveal that their contact networks not only approach isostaticity, but also \emph{minimal isostaticity}, in the thermodynamic limit, therefore partially explaining the  observation in Ref.~\cite{rigidity_loss_three}. We also prove that this is a necessary consequence of purely repulsive interactions in the special case of jamming in circular containers.   
Furthermore, we propose a new design principle based on Assur decomposition for mechanical metamaterials in which the switch of one connection can sharply and remotely control the range of motion and stress propagation.

\emph{Assur decomposition and minimal isostatic graphs (MIGs).---}
We start by reviewing the notion of generic (i.e., combinatorial) rigidity. The mechanical properties of a  frame of vertices (point masses) and edges (springs) denoted as $(G,\mathbf{X})$, come from both its underlying graph $G$ and its  geometry in terms of positions $\mathbf{X}$ of the vertices. Generic rigidity only depends on $G$. A graph $G$ is said to be  rigid if it can be realized into a frame $(G,\mathbf{X})$ that is infinitesimally rigid (i.e., no nontrivial ZMs in the linear theory). Geometries $\mathbf{X}$ that make a generically rigid graph $G$ nonrigid are called geometric singularities. An isostatic graph is a rigid graph in which the removal of any edge leads to a generic ZM (i.e., a ZM in generic realizations). 

An MIG is an isostatic graph with no proper rigid subgraphs~\cite{assur1952issledovanie,DigraphsDecomp,pinned_pebble_game,uniqueenginprops,PENNE199437}. 
The removal of any portion of a MIG results in a global loss of rigidity (otherwise the part that remained rigid would be a proper rigid subgraph). 
So far all studies of MIGs are either on pinned or unpinned  graphs on the plane. 
Fig.~\ref{fig:AssurDecomposition}a shows an example of a pinned isostatic graph, and we
orient the graph, assign direction to each edge, by using the  pebble game algorithm (PGA) for pinned graphs~\cite{pinned_pebble_game} based on Laman's theorem~\cite{laman1970graphs}. 
Here all internal vertices have in-degree of 2, representing the 2 degrees of freedom (DOFs, i.e., the 2 ``pebbles'') of that vertex being constrained 
by the corresponding oriented edges. The ground has in-degree 0 (no inherent DOFs). 
Such a result of the PGA where all edges are oriented and all internal vertices have in-degree 2 indicates that all the DOFs are paired with corresponding constraints and there are no redundant edges. 
The existence of this orientation is sufficient and necessary for pinned isostaticity and the orientation is unique up to the reversal of cycles~\cite{DigraphsDecomp}. Important information is carried by this 2-in orientation: upon the removal of one edge the motion of each vertex is determined by its two immediate ancestors (vertices upstream), and this information is then passed  to its descendants (vertices downstream), as shown in  Fig.~\ref{fig:AssurDecomposition}c,d.  Conversely, stress (i.e., violation of length constraints from the edges) travels upstream as shown in Fig.~\ref{fig:AssurDecomposition}e,f.

The orientation of a pinned isostatic graph  gives a decomposition of this graph into strongly connected components (defined as clusters in which every vertex can be reached from every other vertex along directed edges): the so-called Assur components~\cite{pinned_pebble_game,DigraphsDecomp}.
This decomposition identifies \emph{clusters on the graph where their motion or stress must emerge or disappear in a synchronous way}. Crucially, this decomposition tells us all possible pinned isostatic subgraphs. This is because Assur components of an isostatic graph have a partial order according to the orientation such as depicted in Fig.~\ref{fig:AssurDecomposition}b. Isostatic subgraphs correspond to subsets of these Assur components such that 
for every component all of its ancestors are also in the subset.

This decomposition provides a lower block triangular form for the compatibility matrix $C$, which is defined as the linear map from vertex displacements $u$ to edge extensions $e$, such that $e=C\cdot u$. 
When the rows and columns of the compatibility matrix are arranged according to this partial order, it becomes lower block triangular,
\begin{equation}
    	C=\left( 
    	\begin{matrix}
        	 C^{(1)} & 0 &\cdots & 0  \\
        	 C^{(2,1)} & C^{(2)} &\cdots & 0  \\
          \vdots &\vdots & \ddots & \vdots\\
        	C^{(m,1)} &C^{(m,2)} &\cdots  &C^{(m)}  \\
    	\end{matrix}
    	\right) ,
\end{equation}
where each block in the diagonal is square and full rank and identified with one Assur component.  Mechanically, this tells us that when an edge in component $i$ is pruned, the resulting ZM will involve all vertices in components $j\ge i$, and the addition of any new edge in component $i$ (either between nodes in $i$ or connecting them to the ground) will introduce an SSS that involves all edges in components $j\le i$ (because the equilibrium matrix
$Q=C^T$, is upper block triangular). 

Therefore, for a pinned isostatic graph, (i) it contains no proper isostatic subgraphs, (ii) it is indecomposable (i.e. strongly connected for any 2-in directed orientation), (iii) the compatibility matrix has no proper block triangular decomposition, (iv) removal of any edge results in a ZM that moves all vertices, and (v) forces exerted on any vertex stresses all edges, are all equivalent statements, and graphs satisfying these conditions are called Assur graphs or MIGs~\cite{DigraphsDecomp,uniqueenginprops}.

We next generalize these concepts to frames on the torus, as PBC is adopted in most studies of jamming.  In this case, the 2 trivial translation ZMs always remain, and Laman's theorem for isostaticity has been generalized in Ref.~\cite{Rossthesis} such that 
for a graph of $N$ vertices to be generically isostatic when embedded on the torus (``torus isotatic'' for short), it must be both $(2,2)$-tight [i.e., it has $2N-2$ edges and $(2,2)$-sparse (no subgraph with $n$ vertices has more than $2n-2$ edges)], and all $(2,2)$-tight subgraphs wrap around the torus (i.e., are embedded constructively, following the notion of Ref.~\cite{Rossthesis}).  
We  use the torus PGA algorithm devised in Ref.~\cite{Rossthesis} to obtain orientations of our graphs and from them Assur decompositions (Fig.~\ref{fig:AssurDecomposition}gh).

We thereby generalize the definition of MIGs to torus isostatic graphs, along with their unique  properties in the following theorem [see the Supplementary Material (SM) for the proof].
\begin{theorem}
	Given a graph $G$ that is isostatic on the torus, the following are equivalent
    \begin{enumerate}[label=\alph*]
        \item $G$ has no proper subgraphs that are isostatic on the torus.
    	\item Any orientation of $G$ where the in degree of all but one vertex is  2 is strongly connected on all but that one vertex.
    	\item Removal of any edge results in a generic ZM that moves all vertices relative to each other.
    	\item A generic torque on any edge stresses all bonds
    	\item Adding one edge introduces an SSS that either stresses all edges or  only a non-constructive subgraph.
    		\end{enumerate}
    		\label{th:assurpbc}
\end{theorem}

\begin{figure}
    \centering
    \includegraphics[width=0.5 \textwidth]{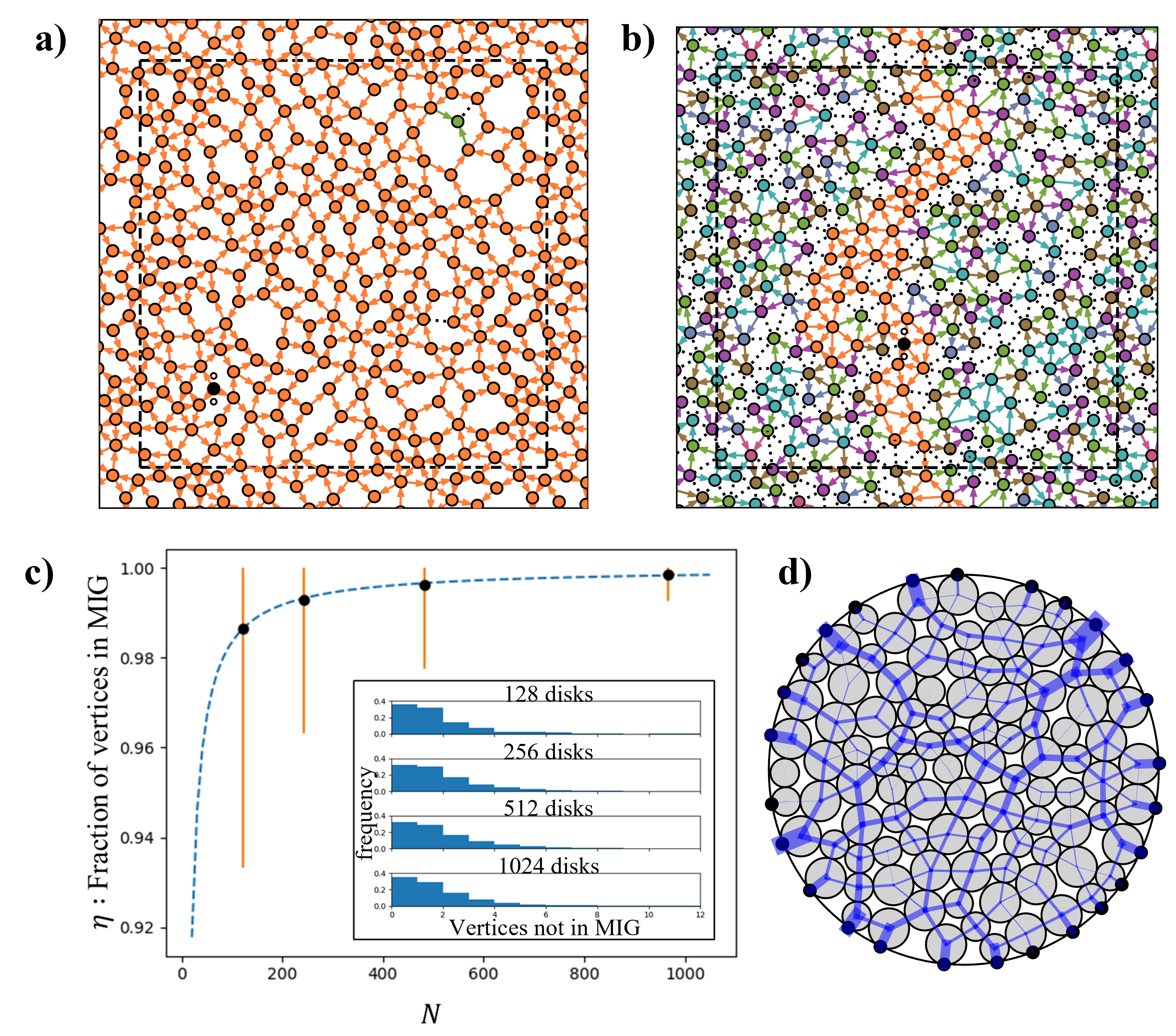}
    \caption{Torus Assur decomposition of jammed packings. (a) An MJP contains a large MIG with almost all vertices. There are two redundant edges in this contact network (black dashed line). (b) A torus isostic graph obtained by pruning redundant edges (black dashed lines) from a packing above the jamming transition contains a large number of Assur components. (Same convention for PBC and ground  as in Fig.~\ref{fig:AssurDecomposition}gh.) (c) The fraction $\eta$ of vertices belonging to the MIG, as a function of total number of vertices in the graph.  The removal of rattlers led to horizontal error bars which are too small to be visible. Vertical error bars indicate 2nd and 98th percentile, from the statistics of testing $100$ packings at each system size, and different choices of the 2 uncovered edges in each packing.
    The inset shows the statistics of the number of vertices not in the MIG, which appears independent of system size. (d) A jammed packing (close to marginal) in a circular container, with a full SSS shown.    
    }
    \label{fig:MJP}
\end{figure}

\emph{Assur decomposition of marginally jammed packings.---} 
We first prepare jammed packings of soft frictionless disks of one-sided Hertzian repulsion by starting from a random configuration with high volume fraction ($\phi=0.90$), and decompress until we have only 2 excess contacts. At each step the energy is minimized using the FIRE algorithm.  For a system of $N$
disks, isostaticity on a torus is reached when the system has $2N-2$ contacts (leaving the 2 trivial translations). This exact state is very hard to reach in large systems, so we choose to terminate decompression at 2 SSSs ($2N$
contacts) and obtain MJPs at various system sizes.  We then run the torus PGA~\cite{Rossthesis} on the contact network, which leaves 2 uncovered edges (i.e., the redundant contacts) and 2 free pebbles (i.e., the 2 trivial ZMs) on the torus. The covered portion of the contact network, which is now directed, is spanning and  torus isostatic.

We then find the Assur components of this spanning isostatic graph, using the strongly connected decomposition adapted for graphs on torus as detailed in the SM.  
This generally results in 
a first Assur component, an MIG by definition from Th.~\ref{th:assurpbc}, covering the entire graph $G$ except a few ``diads'' (an Assur component of 1 vertex and 2 edges), as shown in Fig.~\ref{fig:MJP}a.  We perform this analysis at different system sizes, and collect the fraction $\eta$ of vertices in this MIG (Fig.~\ref{fig:MJP}c).  We find that $\eta=1-\mathcal{O}(N^{-1})$ as $N\to\infty$, meaning that the number of vertices in $G$ not belonging to the MIG does not grow with system size.  
The choice of the ground does not change the Assur decomposition in the case of the MJPs, as we show in the SM.

The fact that almost all nodes belong to the one MIG is a unique property of MJPs. This can be shown by taking other isostatic graphs, e.g., dense packings above the jamming transitition and randomly pruning redundent edges in the contact network until isostaticity is reached.  
This results in a large number of small Assur components (Fig.~\ref{fig:MJP}b), in direct contrast with the case of the MJPs.

\emph{Relation between minimal isostaticity and purely repulsive interactions.---} It is natural to ask whether the minimal isostaticity of MJPs come from the fact that the disks assemble under purely repulsive interactions. 
Interestingly we have an example where this is indeed the case. Consider a set of repulsive disks in a hard frictionles circular container (Fig.~\ref{fig:MJP}d). Contacts of disks with the wall can be represented as edges to the ground, making the contact network a pinned frame. 
A system of fully repulsive disks is only stable if there is an SSS that involves all disks and is compressive at all contacts. 
At the same time, any packing in a circular container has a ZM that involves all disks: a global rotation. Therefore any MJP in a circular container has a full SSS and a full ZM. 
The existence of a realization that satisfies this condition is sufficient and necessary for the graph to be a pinned MIG~\cite{GeometryAssur}. Thus, minimal isostaticity emerges as a consequence of the repulsive nature of the interaction in this case. 
One may conjecture that in the thermodynamic limit, the difference between circular container and torus becomes unimportant, thereby extending the conclusion to the case of PBC. However we don't have a rigorous proof of this argument.

\emph{Relation to observations of global response in Ref.~\cite{rigidity_loss_three}.---} Our findings here partially explain the observation that the removal of any contact makes all vertices in the contact network hinges, and the introduction of any new edge stresses all contacts. This closely relates to Th.~\ref{th:assurpbc}$c$ and $e$. The key difference  is the existence of non-constructive plane isostatic subgraphs [also known as Laman or (2,3)-tight]  with number of vertices $N_{ISG}>3$ (the cases of $N_{ISG}=2$ or $3$ are simply edges and triangles, which don't affect vertices becoming hinges and can't be stressed by the addition of a new edge). If there were no such subgraphs and the network was a MIG on the torus, observations in Ref.~\cite{rigidity_loss_three} follow. 
However if an MIG on the torus had such subgraphs, removal of any edge that is not in such a non-constructive plane isostatic subgraph would make this subgraph, along with all other vertices, mobile, but internal vertices in this subgraph would not be hinges, and adding one edge to this subgraph would only stress this subgraph.
This is a generalization of the minimal isostaticity discussed by Penne~\cite{PENNE199437} where the graph in question is isostatic on the plane and the only allowed plane isostatic graphs
are single bonds. Thus 
results in Ref.~\cite{rigidity_loss_three} indicates that it is very rare to find any  large plane isostatic subgraphs in MJPs, making these networks akin to Penne's MIGs.

\begin{figure}
    \centering
    \includegraphics[width=0.4 \textwidth]{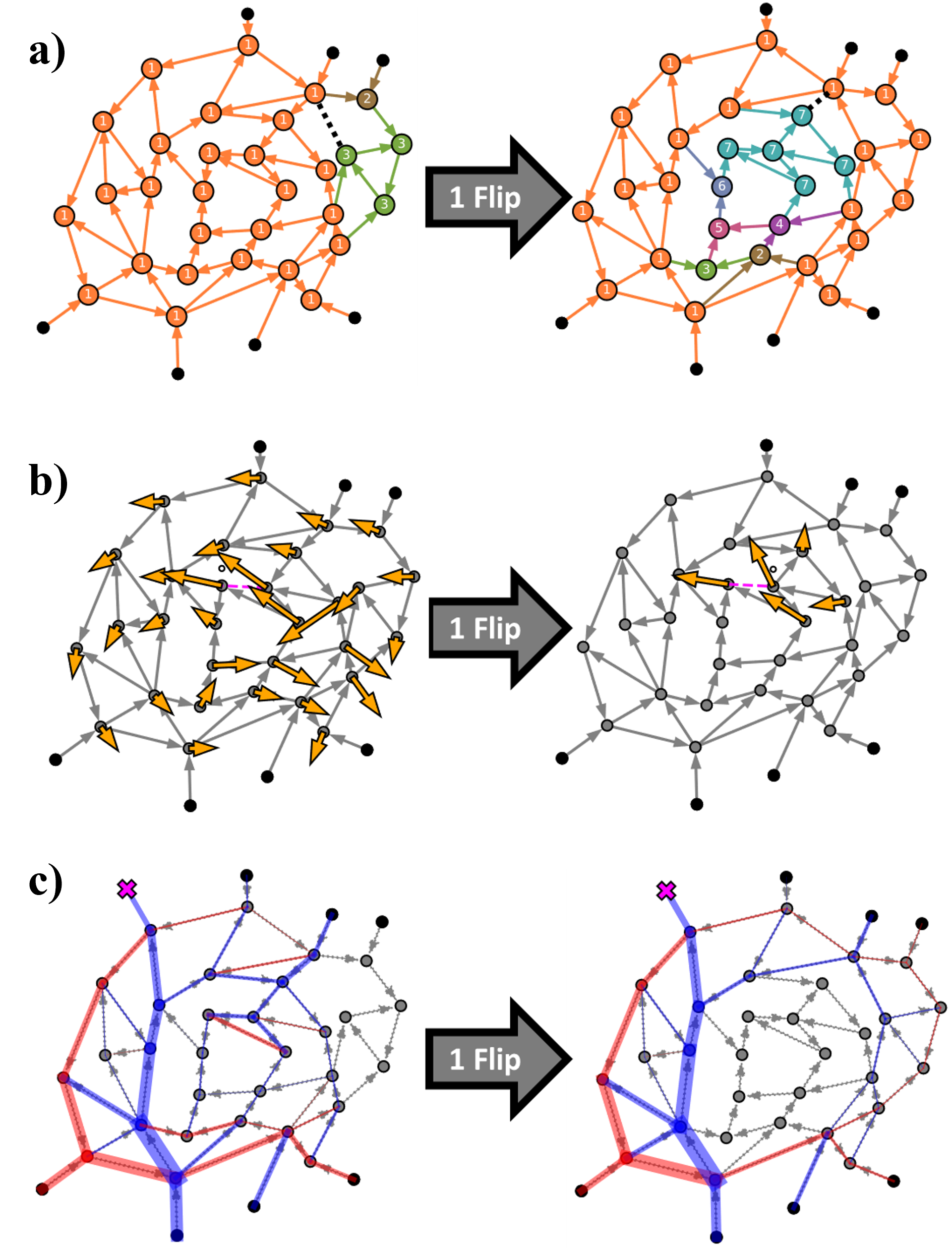}
    \caption{Assur-decomposition based design principle for reconfigurable mechanical metamaterials. (a) Given an isostatic graph, the flip of one edge (black dashed lines) leads to global changes in the Assur Decomposition. (b) The ZMs produced by the pruning of the same edge (magenta dashed line) before and after the flip. (c) The difference in stress response when an external force (marked by the magenta ``X'') is added on the same vertex before and after the flip.
    }
    \label{fig:design}
\end{figure}

\emph{Reconfigurable mechanical metamaterials based on Assur graphs.---} 
The notion of Assur decomposition depicts the remarkable nonlocality of graph rigidity, where a small change in connectivity can affect rigidity \emph{arbitrarily far away}~\cite{PG_Original}.  This unique property can be powerful in the design of 
``Maxwell'' mechanical metamaterials where the proximity to isostaticity gives rise to rich phenomena in terms of modes, stress, and reconfiguration~\cite{paulose2015topological,rocklin2017transformable,rocklin2018folding,Sun2020,Xiu2022,Zhang2022}.

Here we propose to utilize Assur graphs to design mechanical metamaterials which reconfigure the spatial distribution of motion and stress, and thus precisely direct actuations.
One such example is shown in Fig.~\ref{fig:design}, where the flip of just one edge changes the Assur decomposition of the whole graph, because we loop a far-downstream vertex back to the MIG.  This causes the whole interior region of the graph to change from downstream Assur components to  part of the MIG. Thus, before the flip the interior region can be mobile (upon pruning any edges in the region) without moving the MIG and will not be stressed when the MIG is stressed, whereas in after the flip their motion and stress are locked in the MIG.  
A unique advantage of this principle is that the graph remain isostatic before the after the reconfiguration, maintaining good stability.

\emph{Conclusions and discussions.---} Here we examine substructures of isostatic graphs in physical system, by extending the concept of Assur decomposition to graphs with PBC and show that MJPs are not only isostatic but minimally isostatic, as well as proposing a new design rule for reconfigurable mechanical metamaterials reconfigurable via remote mechanical control.  

Many intriguing new questions follow from these findings: What is the substructure of rigidity in packings of more complex particles such as frictional~\cite{bi2011jamming},  nonspherical particles~\cite{Mailman2009}, or packings in 3D~\cite{OHern2003}? How to optimize this type of mechanical remote control to obtain significant change in motion and stress propagation in experimental systems with imperfections? How to obtain networks with mechanical properties that resemble jammed packings without the packing process~\cite{Lopez2013,hagh2018jamming}?
And more interestingly, can we control isostatic substructures by programming particle properties and protocols, thereby obtaining self-assembled reconfigurable metamaterials? These would be interesting questions for future research.

\emph{Acknowledgements.---} This work was supported in part by the National Science Foundation under Grant No.~NSF-DMR-2026825 and the Office of Naval Research under Grant No.~MURI N00014-20-1-2479.

\appendix


\section{Theorems about rigidity on torus}
Here we first list two theorems from the literature on rigidity, (2,2)-tight graphs, and orientation.

\begin{theorem}\label{TH:GeneralizedLaman}
    A graph $G$ with its embedding on the torus is isostatic on the torus iff it is (2,2)-tight and all of its (2,2)-tight sub-graphs are embedded constructively.
    \label{th:lamman22}
\end{theorem}
This theorem is the generalization of Laman's theorem to graphs on the torus, and was proven in Ref.~\cite{Rossthesis} (Theorem 6.3.1).  
Note that the language we use is slightly different but equivalent to the version in Ref.~\cite{Rossthesis}.  
We have written it in this way since it explicitly states the condition checked by the pebble game on the torus. The main difference between the two statements of the theorem is that Ref.~\cite{Rossthesis} encodes the information of the embedding in a gain assignment. When we say ``all of its (2,2)-tight sub-graphs are embedded constructively" it is equivalent to saying that the ``gain assignment is constructive" in the language of Ref.~\cite{Rossthesis}.

\begin{theorem}
    A graph $G$ is (2,2)-tight iff there exist an orientation such that one vertex has in-degree 0 and all others have in degree 2.
    \label{lemma:orientationtight}
\end{theorem}

This theorem follows from Theorem 8 in Ref.~\cite{pga_sparse_graphs}, where it is shown that  
a graph is (2,2)-tight iff it can be fully oriented by the (2,2)-pebble game with only two free pebbles remaining. We can place the two free pebbles on any vertex, then it would have in-degree 0 and the rest of the vertices would have in-degree 2. These pebble game orientations are all the possible orientations such that the in-degree of one vertex is 0 and the rest have in-degree 2. This follows from the fact the once the in-degree of each vertex has been specified the orientation is unique up to reversal of cycles~\cite{DigraphsDecomp}. 

\section{Proof of Theorem 1 in the main text}
\subsection{Th.1.a $\Longleftrightarrow$Th.1.b}
We will show that statement \textit{Th.1.a: ``$G$ has no proper subgraphs that are isostatic on the torus"} is equivalent to statement \textit{Th.1.b ``Any orientation of $G$ where the in-degree of all but one vertex is 2 is strongly connected on all but that one vertex."}, given that $G$ is isostatic on the torus.  

Let us define ``minimally (2,2)-tight graphs'' as graphs that are (2,2)-tight and has no proper (2,2)-tight subgraphs. Then from Th.~\ref{TH:GeneralizedLaman} it follows immediately:

\begin{lemma}
    A graph $G$ is minimally isostatic on the torus iff it is minimally (2,2)-tight and is embedded constructively.
    \label{lemma:m22t}
\end{lemma}

Next we prove a lemma that relates ``minimally (2,2)-tight'' with graph orientations.

\begin{lemma}\label{lemma:minimalorientation}
    A graph $G$ is minimally (2,2)-tight iff it is (2,2)-tight and all orientations in which one vertex has in-degree 0 and all others have in-degree 2 are strongly connected on the in-degree 2 vertices.
\end{lemma}

\emph{Proof.}  
Consider a (2,2)-tight graph $G$.  
We  prove this lemma by contradiction.  
Suppose there exist a proper subgraph $G'$ that is (2,2)-tight,
the number of edges in $G'$ is $N'_B=2N'-2$ where $N'$ is the number of vertices in $G'$.  Because the 0 in-degree vertex can be chosen to be any vertex,(See \cite{pga_sparse_graphs} specifically Lemma 13) we can place it in $G'$.  Thus, the in-degree of the subgraph $G'$ is given by $InDegree(G') = 2(N'-1)-N_B'=0$.  
This means that there must be no edges pointing from the rest of $G$ to $G'$, and thus $G'$ is a ``source'' for the rest of $G$.  Then $G$ is not strongly connected on all in-degree 2 vertices.  Thus, if  all orientations such that one vertex has in-degree 0 and all others have in degree 2 are strongly connected on the in degree 2 vertices, $G$ is  minimally (2,2)-tight.

Conversely, if $G$ is (2,2)-tight but there exist an orientation that is not strongly connected on all in-degree 2 vertices, there must be a ``source'' in which one vertex has in-degree 0 and all others have in-degree 2.  As a result, by Th.~\ref{lemma:orientationtight}, this ``source'' is a proper (2,2)-tight subgraph, and thus $G$ is not minimally (2,2)-tight.  Therefore, $G$ being  minimally (2,2)-tight indicates that all all orientations in which one vertex has in-degree 0 and all others have in degree 2 are strongly connected on the in degree 2 vertices.  These two arguments together prove Lemma~\ref{lemma:minimalorientation}

The combination of Lemma~\ref{lemma:m22t} and Lemma~\ref{lemma:minimalorientation} imply \textit{Th.1.a} $\Longleftrightarrow$\textit{Th.1.b}.

\subsection{Th.1.a $\Longleftrightarrow$Th.1.c}
This is rather straightforward, given $G$ being minimally isostatic on the torus the loss of an edge results in a floppy mode. The part of the graph that does not become generically floppy as a result of this edge being pruned will be a proper rigid subgraph. No such subgraphs exist (since $G$ is minimally isostatic) therefore all the vertices become floppy. To prove the converse note that if a graph become fully floppy when one edge is cut it means that any rigid subgraph must have contained that edge. If the graph becomes fully floppy upon the cutting of any edge then all any rigid subgraph must contain all edges, therefore there are no proper rigid subgraphs. \\

\subsection{Th.1.a $\Longleftrightarrow$Th.1.d,e}
Now let us move on to the equivalence of Th.1.a $\Longleftrightarrow$Th.1.d and Th.1.a $\Longleftrightarrow$Th.1.e, 
the stress conditions. 

Rigidity in general requires that  given any external force $\mathbf{f}$ orthogonal to the trivial degrees of freedom there can be found a set of bond tensions $\mathbf{t}$ that balances this external force $\mathbf{f}=Q\mathbf{t}$. We will call such a set of tensions the response. It is clear that for isostatic networks the response must be unique otherwise if we had distinct responses  $\mathbf{t}_1,\mathbf{t}_2$ we could find a state of self stress $0=Q(\mathbf{t}_1-\mathbf{t}_2)$. Given the uniqueness of the response it is then clear that for isostatic networks the response to an external force $\mathbf{f}$ will be provided only by the smallest subgraph capable of providing a response. The equivalence of 1.a to 1.d and 1.e follow immediately from this discussion. A subgraph can provide a response iff the force is orthogonal to its space of zero modes.  Note that while only a torus isostatic subgraph can respond to a torque, if we exert a force that is orthogonal to translations and rotation (e.g., a force dipole induced by the addition of an extra edge, as stated in Th.1.e) then a nonconstructive plane isostatic (i.e Laman) subgraph can respond.

\subsection{Generalization to Minimally (k,k)-tight graphs}
We can generalize Lemma \ref{lemma:minimalorientation} to minimally $(k,k)$ tight graphs since the proof for arbitrary $k$ is completely analogous to the $k=2$ case which we have done.
\begin{theorem}
    A graph $G$ is minimally $(k,k)$-tight iff it is $(k,k)$-tight and all orientations such that one vertex has in-degree 0 and all others have in-degree $k$ are strongly connected on the in-degree $k$ vertices.
\end{theorem}

The effects of pruning an edge or adding an external edge are also generalized. 
It is an interesting question to find the physical setup that allows us to realize a minimally $(k,k)$-tight graph as a mechanical frame. The (3,3)-tight case is achieved by embedding the graph on the plane, without pins, and having the vertices be rigid bodies in 2D which therefore would have 3 inherent degrees of freedom. 
Another interesting physical example are point-like particles in 3D under periodic boundary conditions, which leads to 
isostatic graphs on  3D torus.  However it is worth noting that (3,3)-tightness plus constructive embedding will most likely be a necessary but not sufficient condition for isostaticity on the 3D torus.\cite{3DPGA}\\

\section{Assur Decomposition on the Torus and Its Uniqueness}

\begin{figure}
    \centering
    \includegraphics[width=0.30\textwidth]{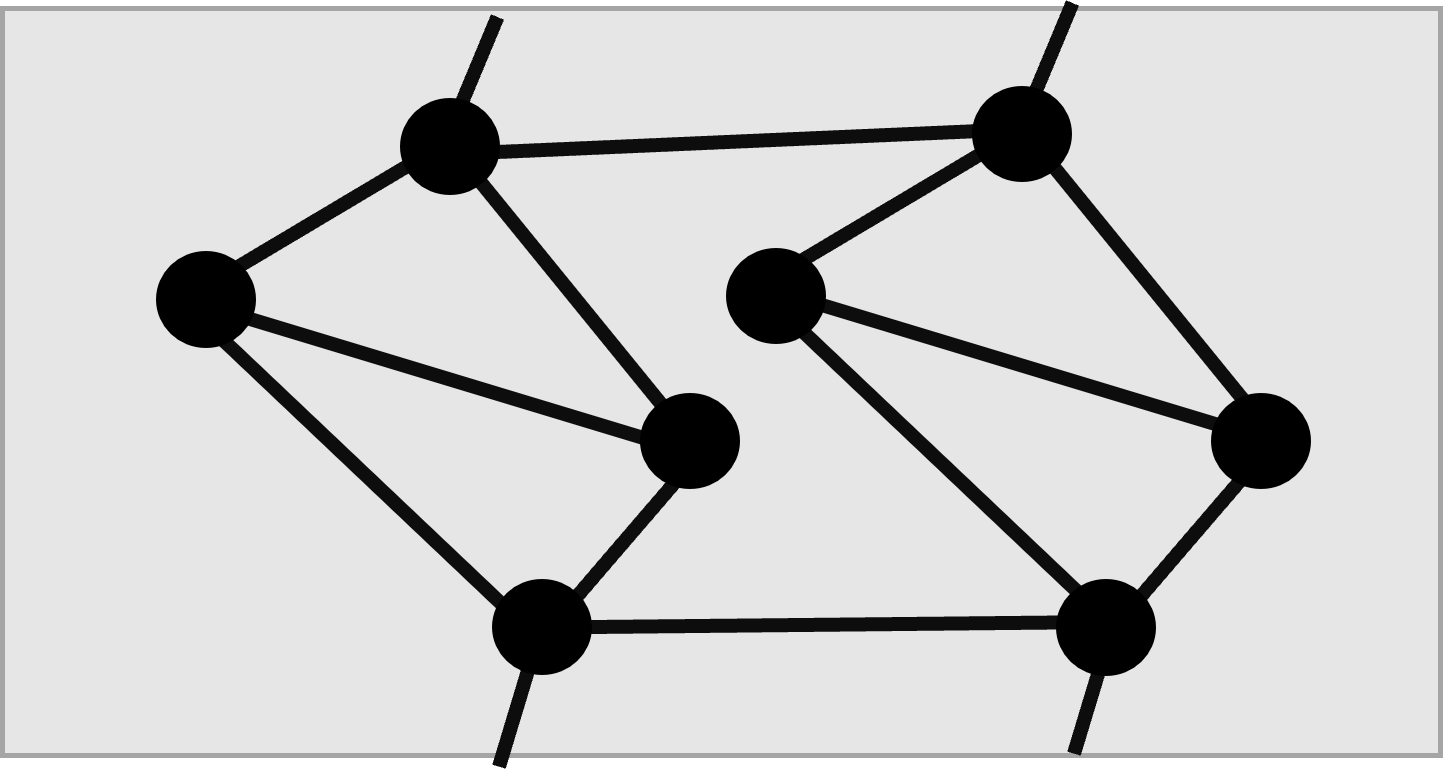}
   \caption{An example of a graph, isostatic on the torus with two distinct minimally isostatic subgraphs and therefore a non unique Assur decomposition.}
    \label{fig:nonu}
\end{figure}

Any graph that is isostatic on the torus has at least one subgraph that is minimally isostatic on the torus, although it may have multiple as in Fig.~\ref{fig:nonu}.  If there exist multiple disjoint minimally isostatic subgraphs our generalization of the Assur decomposition will not be unique as in the pinned case. 

Fortunately, this is not an issue for  isostatic graphs obtained from MJPs. This networks always had a unique minimally isostatic subgraph. In what follows we explain the Assur decomposition on the torus and why, given that there are not multiple minimally isostatic subgraphs the decomposition we propose is unique.

What would make the Assur decomposition non-unique when there are multiple disjoint minimally isostatic subgraphs is
choice of ground $s_0$, which is the vertex chosen to have in-degree 0. For any (2,2)-tight graph given the choice of ground the decomposition into strongly connected components is unique. Note, $s_0$ will always be its own strongly connected component. Consider a vertex $s_1$ which is a neighbor of $s_0$, which is to say the edge $(s_0,s_1)$ belongs to the graph. The strongly connected component to which $s_1$ belongs to, along with the ground $s_0$ constitute what is called the critical subgraph of edge $(s_0,s_1)$. The critical subgraph is the smallest (2,2)-tight subgraph that contains $(s_0,s_1)$ (see Lemma 2.6.4d in Ref.~\cite{Rossthesis}). It is clear that the critical subgraph is always strongly connected on all vertices except the ground, but this does not imply that this subgraph is minimally (2,2)-tight. Recall Lemma \ref{lemma:minimalorientation} requires all possible orientations to be strongly connected on the non-ground sites for the graph to be minimally (2,2)-tight, and the existence of a particular orientation that is strongly connected is not sufficient. If we want to identify the critical subgraph $(s_0,s_1)$ as a first Assur component it must be indecomposable for any choice of ground, which is to say minimally (2,2)-tight. In order to do this we must choose a ground that belongs to a minimally (2,2)-tight subgraph. Luckily the PGA on the torus (Ref.~\cite{Rossthesis}) naturally does this as we discuss below. 

The first step of the PGA on the torus is to orient according to the (2,3)-PGA. Then in the second step the bonds that remain uncovered are checked. Three pebbles are called to such bonds, if 3 pebbles are found and the critical subgraph of the bond is constructive then the bond is covered. Note, just after the first step the subgraph composed of all covered bonds is (2,3)-tight and therefore contains no (2,2)-tight subgraphs. When the first bond $(s_0,s_1)$ is covered during the second step the covered subgraph obtains a (2,2)-tight subgraph. This implies that all (2,2)-tight subgraphs of the covered subgraph contain the edge  $(s_0,s_1)$. Therefore the first bond covered during this second step belongs to a minimally-(2,2) tight graph which will be identical to its critical subgraph. For a graph rigid on the torus with no disjoint minimally isotatic subgraphs only one bond will be covered during the second step. This is what we found to always happen for MJP contact networks. Then we see that in such cases choosing the ground to be one of the vertices of the last covered bond guarantees that the critical subgraph is minimally (2,2)-tight and therefore we can identify is as the first Assur component. Specifically, the steps of our decomposition are as follows

Given $G$ isostatic on the torus:
\begin{enumerate}
    \item Orient $G$ according to the PGA on the torus
    \item Place the remaining two pebbles on one of the ends of the last covered edge, making it the ground $s_0$.  (This happens automatically when the PGA finishes.)
    \item Partition the vertices into strongly connected components according the PGA orientation
    \item Define an assur component for each strongly connected component except the ground as follows: All the vertices in the strongly connected component along with edges pointing to these vertices.
    \item Choose a neighbor $s_1$ of the ground vertex $s_0$, assign the ground $s_0$ to the same Assur component as $s_1$ is in. This is an induced subgraph of $G$ and also minimally isostatic on the torus.
\end{enumerate}

\begin{figure}
    \centering
    \includegraphics[width=0.30\textwidth]{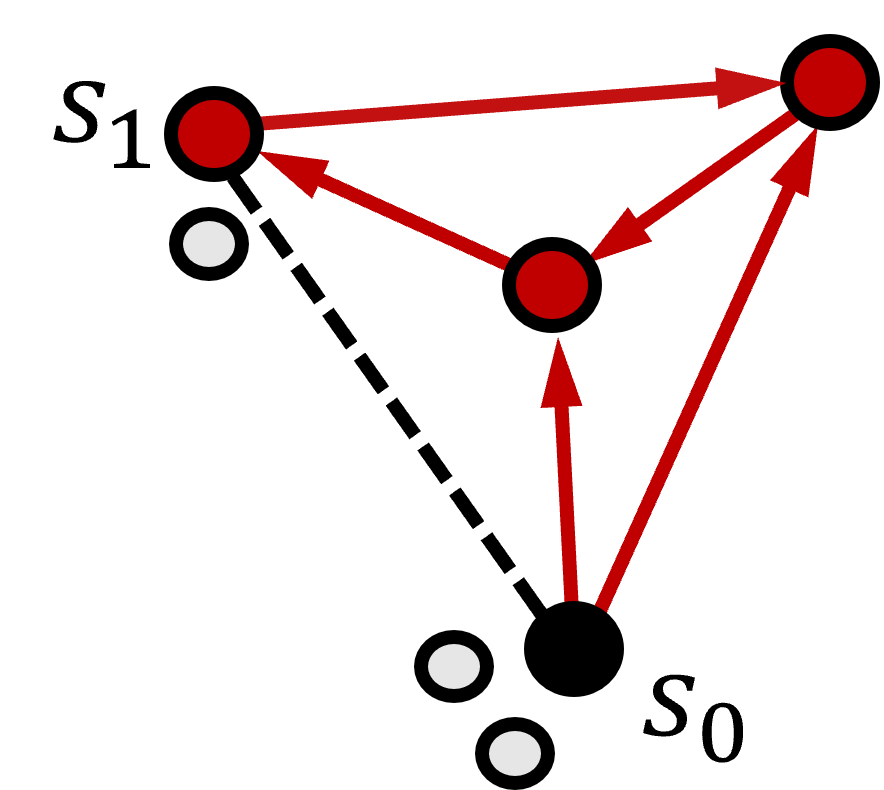}
    \caption{A minimally (2,2)-tight graph oriented according to the pebble game on the plane. Note that all vertices except $s_0$ are strongly connected.}
    \label{fig:k4pg}
\end{figure}

It is perhaps worth noting that every  minimally (2,2)-tight graph can be realized as a MIG on the torus, a pinned MIG (i.e., an Assur graph), 
and as a free frame on the plane with only one SSS (which must be a global SSS), sometimes called a Laman circuit.
And so we see the equivalence between these three structures. To see that frames on the plane with only one SSS, which is global, are always minimally (2,2)-tight, note that when oriented according to the (2,3)-PGA as in Fig.~\ref{fig:k4pg} the critical subgraph always identifies a subgraph with one global SSS. This is a result that goes back to Jacobs and Thorpe \cite{PG_Original}. It is interesting that this full stress condition, satisfied by MJPs, shows up here. This is related to the special geometric singularity of pinned Assur graphs which we mention in the main text and leads to MJPs on circular containers to necessarily have Assur graphs as their contact networks. Also, note that a 2D MJP on the surface of a sphere will necessarily be a Laman circuit and therefore minimally (2,2)-tight. Such simple arguments don't seem to extend straightforwardly to the torus and so the question of how the fully repulsive interaction leads in general to minimal isostaticity remains an open problem.


\begin{thebibliography}{10}

\bibitem{Zhang2019}
S.~Zhang, L.~Zhang, M.~Bouzid, D.~Z. Rocklin, E.~Del~Gado, and X.~Mao,
  ``Correlated rigidity percolation and colloidal gels,'' {\em Phys. Rev.
  Lett.}, vol.~123, p.~058001, Jul 2019.

\bibitem{colombo2014stress}
J.~Colombo and E.~Del~Gado, ``Stress localization, stiffening, and yielding in
  a model colloidal gel,'' {\em Journal of rheology}, vol.~58, no.~5,
  pp.~1089--1116, 2014.

\bibitem{OHern2003}
C.~S. O'Hern, L.~E. Silbert, A.~J. Liu, and S.~R. Nagel, ``Jamming at zero
  temperature and zero applied stress: The epitome of disorder,'' {\em Phys.
  Rev. E}, vol.~68, p.~011306, Jul 2003.

\bibitem{bi2015density}
D.~Bi, J.~Lopez, J.~M. Schwarz, and M.~L. Manning, ``A density-independent
  rigidity transition in biological tissues,'' {\em Nature Physics}, vol.~11,
  no.~12, pp.~1074--1079, 2015.

\bibitem{Lubensky2015}
T.~C. Lubensky, C.~L. Kane, X.~Mao, A.~Souslov, and K.~Sun, ``Phonons and
  elasticity in critically coordinated lattices,'' {\em Reports on Progress in
  Physics}, vol.~78, p.~073901, jun 2015.

\bibitem{Mao2018}
X.~Mao and T.~C. Lubensky, ``Maxwell lattices and topological mechanics,'' {\em
  Annual Review of Condensed Matter Physics}, vol.~9, no.~1, pp.~413--433,
  2018.

\bibitem{laman1970graphs}
G.~Laman, ``On graphs and rigidity of plane skeletal structures,'' {\em Journal
  of Engineering mathematics}, vol.~4, no.~4, pp.~331--340, 1970.

\bibitem{asimow1978rigidity}
L.~Asimow and B.~Roth, ``The rigidity of graphs,'' {\em Transactions of the
  American Mathematical Society}, vol.~245, pp.~279--289, 1978.

\bibitem{alexander1998amorphous}
S.~Alexander, ``Amorphous solids: their structure, lattice dynamics and
  elasticity,'' {\em Physics reports}, vol.~296, no.~2-4, pp.~65--236, 1998.

\bibitem{van2009jamming}
M.~van Hecke, ``Jamming of soft particles: geometry, mechanics, scaling and
  isostaticity,'' {\em Journal of Physics: Condensed Matter}, vol.~22, no.~3,
  p.~033101, 2009.

\bibitem{liu2010jamming}
A.~J. Liu and S.~R. Nagel, ``The jamming transition and the marginally jammed
  solid,'' {\em Annu. Rev. Condens. Matter Phys.}, vol.~1, no.~1, pp.~347--369,
  2010.

\bibitem{goodrich2016scaling}
C.~P. Goodrich, A.~J. Liu, and J.~P. Sethna, ``Scaling ansatz for the jamming
  transition,'' {\em Proceedings of the National Academy of Sciences},
  vol.~113, no.~35, pp.~9745--9750, 2016.

\bibitem{jacobs1995generic}
D.~J. Jacobs and M.~F. Thorpe, ``Generic rigidity percolation: the pebble
  game,'' {\em Physical review letters}, vol.~75, no.~22, p.~4051, 1995.

\bibitem{donev2004linear}
A.~Donev, S.~Torquato, F.~H. Stillinger, and R.~Connelly, ``A linear
  programming algorithm to test for jamming in hard-sphere packings,'' {\em
  Journal of Computational Physics}, vol.~197, no.~1, pp.~139--166, 2004.

\bibitem{Mailman2009}
M.~Mailman, C.~F. Schreck, C.~S. O'Hern, and B.~Chakraborty, ``Jamming in
  systems composed of frictionless ellipse-shaped particles,'' {\em Phys. Rev.
  Lett.}, vol.~102, p.~255501, Jun 2009.

\bibitem{broedersz2011criticality}
C.~P. Broedersz, X.~Mao, T.~C. Lubensky, and F.~C. MacKintosh, ``Criticality
  and isostaticity in fibre networks,'' {\em Nature Physics}, vol.~7, no.~12,
  pp.~983--988, 2011.

\bibitem{Ellenbroek_2011}
W.~G. Ellenbroek and X.~Mao, ``Rigidity percolation on the square lattice,''
  {\em Europhysics Letters}, vol.~96, p.~54002, nov 2011.

\bibitem{bi2011jamming}
D.~Bi, J.~Zhang, B.~Chakraborty, and R.~P. Behringer, ``Jamming by shear,''
  {\em Nature}, vol.~480, no.~7377, pp.~355--358, 2011.

\bibitem{Yan2014}
L.~Yan and M.~Wyart, ``Evolution of covalent networks under cooling:
  Contrasting the rigidity window and jamming scenarios,'' {\em Phys. Rev.
  Lett.}, vol.~113, p.~215504, Nov 2014.

\bibitem{rigidity_loss_three}
W.~Ellenbroek, V.~Hagh, A.~Kumar, M.~Thorpe, and M.~Hecke, ``Rigidity loss in
  disordered systems: Three scenarios,'' {\em Physical review letters},
  vol.~114, 11 2014.

\bibitem{Henkes2016}
S.~Henkes, D.~A. Quint, Y.~Fily, and J.~M. Schwarz, ``Rigid cluster
  decomposition reveals criticality in frictional jamming,'' {\em Phys. Rev.
  Lett.}, vol.~116, p.~028301, Jan 2016.

\bibitem{behringer2018physics}
R.~P. Behringer and B.~Chakraborty, ``The physics of jamming for granular
  materials: a review,'' {\em Reports on Progress in Physics}, vol.~82, no.~1,
  p.~012601, 2018.

\bibitem{assur1952issledovanie}
L.~V. Assur and I.~I. Artobolevskij, {\em Issledovanie ploskih ster{\v{z}}nevyh
  mehanizmov s niz{\v{s}}imi parami s to{\v{c}}ki zreni{\^a} ih struktury i
  klassifikacii}.
\newblock Izdatel'stvo Akademii Nauk SSSR, 1952.

\bibitem{DigraphsDecomp}
O.~Shai, A.~Sljoka, and W.~Whiteley, ``Directed graphs, decompositions, and
  spatial linkages,'' {\em Discrete Applied Mathematics}, vol.~161, no.~18,
  pp.~3028 -- 3047, 2013.

\bibitem{pinned_pebble_game}
A.~Sljoka, ``Checking mobility and decomposition of linkages via pebble game
  algorithm,'' 08 2011.

\bibitem{uniqueenginprops}
E.~Hahn and O.~Shai, ``The unique engineering properties of assur
  groups/graphs, assur kinematic chains, baranov trusses and parallel robots,''
  in {\em International Design Engineering Technical Conferences and Computers
  and Information in Engineering Conference}, vol.~50169, p.~V05BT07A074,
  American Society of Mechanical Engineers, 2016.

\bibitem{PENNE199437}
R.~Penne, ``Isostatic bar and joint frameworks in the plane with irreducible
  pure conditions,'' {\em Discrete Applied Mathematics}, vol.~55, no.~1,
  pp.~37--57, 1994.

\bibitem{chen2022modular}
S.~Chen, F.~Giardina, G.~P. Choi, and L.~Mahadevan, ``Modular representation
  and control of floppy networks,'' {\em Proceedings of the Royal Society A},
  vol.~478, no.~2264, p.~20220082, 2022.

\bibitem{Rossthesis}
E.~Ross, {\em Geometric and Combinatorial Rigidity of Periodic Frameworks as
  Graphs on The Torus}.
\newblock PhD thesis, 05 2011.

\bibitem{GeometryAssur}
B.~Servatius, O.~Shai, and W.~Whiteley, ``Geometric properties of assur
  graphs,'' {\em European Journal of Combinatorics}, vol.~31, no.~4,
  pp.~1105--1120, 2010.
\newblock Rigidity and Related Topics in Geometry.

\bibitem{PG_Original}
D.~J. Jacobs and M.~F. Thorpe, ``Generic rigidity percolation: The pebble
  game,'' {\em Phys. Rev. Lett.}, vol.~75, pp.~4051--4054, Nov 1995.

\bibitem{paulose2015topological}
J.~Paulose, B.~G.-g. Chen, and V.~Vitelli, ``Topological modes bound to
  dislocations in mechanical metamaterials,'' {\em Nature Physics}, vol.~11,
  no.~2, pp.~153--156, 2015.

\bibitem{rocklin2017transformable}
D.~Rocklin, S.~Zhou, K.~Sun, and X.~Mao, ``Transformable topological mechanical
  metamaterials,'' {\em Nature communications}, vol.~8, no.~1, pp.~1--9, 2017.

\bibitem{rocklin2018folding}
D.~Rocklin, V.~Vitelli, and X.~Mao, ``Folding mechanisms at finite
  temperature,'' {\em arXiv preprint arXiv:1802.02704}, 2018.

\bibitem{Sun2020}
K.~Sun and X.~Mao, ``Continuum theory for topological edge soft modes,'' {\em
  Phys. Rev. Lett.}, vol.~124, p.~207601, May 2020.

\bibitem{Xiu2022}
H.~Xiu, H.~Liu, A.~Poli, G.~Wan, K.~Sun, E.~M. Arruda, X.~Mao, and Z.~Chen,
  ``Topological transformability and reprogrammability of multistable
  mechanical metamaterials,'' {\em Proceedings of the National Academy of
  Sciences}, vol.~119, no.~52, p.~e2211725119, 2022.

\bibitem{Lopez2013}
J.~H. Lopez, L.~Cao, and J.~M. Schwarz, ``Jamming graphs: A local approach to
  global mechanical rigidity,'' {\em Phys. Rev. E}, vol.~88, p.~062130, Dec
  2013.

\bibitem{hagh2018jamming}
V.~F. Hagh, E.~I. Corwin, K.~Stephenson, and M.~Thorpe, ``Jamming in
  perspective,'' {\em arXiv preprint arXiv:1803.03869}, 2018.

\bibitem{pga_sparse_graphs}
A.~Lee and I.~Streinu, ``Pebble game algorithms and sparse graphs,'' {\em
  Discrete Mathematics}, vol.~308, no.~8, pp.~1425--1437, 2008.
\newblock Third European Conference on Combinatorics.

\bibitem{3DPGA}
M.~V. Chubynsky and M.~F. Thorpe, ``Algorithms for three-dimensional rigidity
  analysis and a first-order percolation transition,'' {\em Phys. Rev. E},
  vol.~76, p.~041135, Oct 2007.

\end{thebibliography}
\bibliographystyle{ieeetr}

\end{document}